\documentclass{DISproc}

\usepackage{hepunits}

\begin{document}
%------------------------------------
\title{Spacelike and Timelike Compton Scattering: \\Progress report}

%for single authors the superscripts are optional
\author{{\slshape H. Moutarde$^1$, F. Sabati\'e$^1$}\\[1ex]
$^1$IRFU/Service de Physique Nucl\'eaire, CEA, Centre de Saclay, F-91191 Gif-sur-Yvette, FRANCE}

% please enter the contribution ID for the DOI
\contribID{154}

\doi  % if there is an online version we will register DOIs

\maketitle

\begin{abstract}
We present some recent results on the analysis of hard scattering processes in the framework of Generalized Parton Distributions. In particular we compute DVCS observables on unpolarized targets with the Kroll~-~Goloskokov model (suited to DVMP analysis). We also discuss NLO contributions to DVCS and TCS processes for various kinematic settings.
\end{abstract}

%------------------%
%--   Introduction   --%
%------------------%

\section{Introduction}

The Deeply Virtual Compton Scattering (DVCS) process is the theoretically cleanest way to access Generalized Parton Distributions (GPD). However Deeply Virtual Meson Production (DVMP) and Timelike Compton Scattering (TCS) measurements will bring further constraints on our experimental knowledge of GPDs (see reviews~\cite{Goeke:2001tz, Diehl:2003ny, Belitsky:2005qn, Boffi:2007yc} and references therein). 

First we outline some results on exclusive processes and describe a GPD model used for the evaluations presented in this work. Then we estimate the phenomenological impact of Next-to-Leading Order (NLO) corrections to Leading Order (LO) evaluations. The following section confront this GPD model to DVCS measurements. We finish with some technical remarks.

%----------------------------------%
%--   1st part : theoretical framework   --%
%----------------------------------%
\section{Theoretical framework}

\subsection{Exclusive processes}

The partonic interpretation of electroproduction of mesons or real photon relies on the use of factorization theorems. They express observables in terms of Compton Form Factors (CFF), which are convolutions of known kernels with GPDs. That GPDs are universal quantities should be checked to ensure the consistency of this partonic picture. One first step towards this aim consists in confronting a GPD model tailored to study DVMP to DVCS.

\subsection{Kroll~-~Goloskokov GPD model}

The Kroll~-~Goloskokov (KG) model was designed to interpret meson electroproduction. Details about this model can be found in \cite{Goloskokov:2005sd, Goloskokov:2008ib, Diehl:2004cx}. The GPD $H$ (main contribution to the DVCS observables discussed here) is classically described by a double distribution and a profile function. It is Regge behaved and possesses an exponential dependence in Mandelstam variable $t$, uncorrelated to the longitudinal momentum transfer $x$. Its corresponding CFF is denoted $\mathcal{H}$.

%-----------------------------------------%
%--   2nd part : DVCS and TCS at LO and NLO   --%
%-----------------------------------------%

\section{DVCS and TCS at LO and NLO}

\subsection{LO and NLO Compton Form Factors}

Since the integration kernel of CFFs is singular in the vicinity of the skewness $\xi$, a CFF a is complex function. At LO the imaginary part of a CFF is simply the singlet GPD evaluated at $x = \xi$, but at NLO both real and imaginary parts involve integrals with logarithmic integrable singularities. Their numerical treatment requires some care, espacially at small $\xi$ \cite{moutarde:2012a}. Expressions for CFFs at LO and NLO for DVCS and TCS may be found in \cite{Belitsky:1998pq, Belitsky:1998uk, Pire:2011st}.

\subsection{Estimates for the DVCS and TCS processes}

\begin{figure}[htb]
  \centering
%  {\rotatebox{-90}{\includegraphics[height=0.7\textwidth]{moutarde_herve_fig1}}}
  \includegraphics[width=0.7\textwidth]{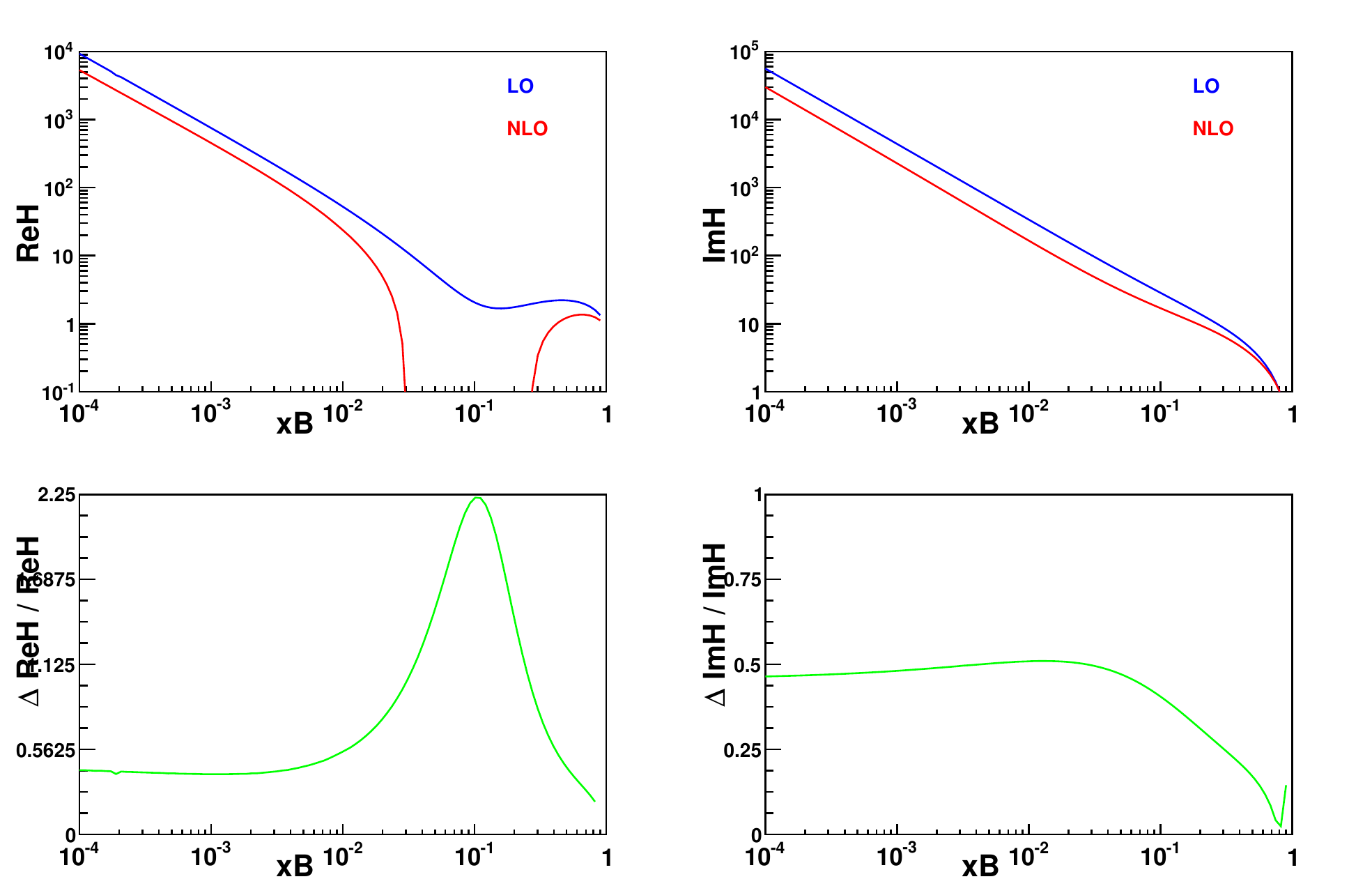}
  \caption{Upper plots: $\mathcal{H}$ at LO and NLO. Lower plots: relative discrepancy at LO and NLO.}
  \label{Fig:CFFLOvsNLO}
\end{figure}

Figure~\ref{Fig:CFFLOvsNLO} displays the real and imaginary parts of the CFF $\mathcal{H}$ at LO and NLO evaluated at factorization scale \unit{4}{\GeV^2} and vanishing $t$ for Bjorken $x_B = 2 \xi / ( 1 + \xi )$ ranging between $10^{-4}$ and 1. Although the comparison is model dependent, the typical discrepancy between LO and NLO is 40~\% at small $\xi$. It is maximum around $\xi = 0.1$ (COMPASS - HERMES kinematics).

%---------------------------------------------%
%--   3rd part : Computation of DVCS observables   --%
%---------------------------------------------%

\section{Computation of DVCS observables}

\subsection{HERMES observables}

\begin{figure}[htb]	
  \centering
  \begin{tabular}{cc}	
    \includegraphics[width=0.35\textwidth]{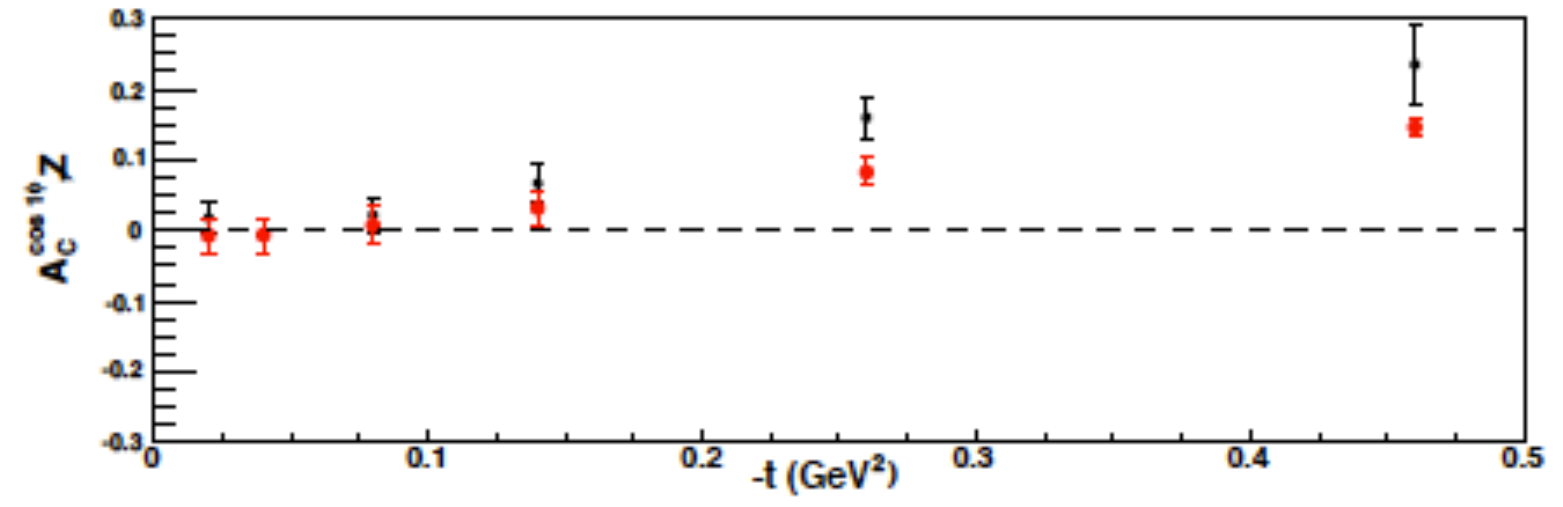}
     &
    \includegraphics[width=0.35\textwidth]{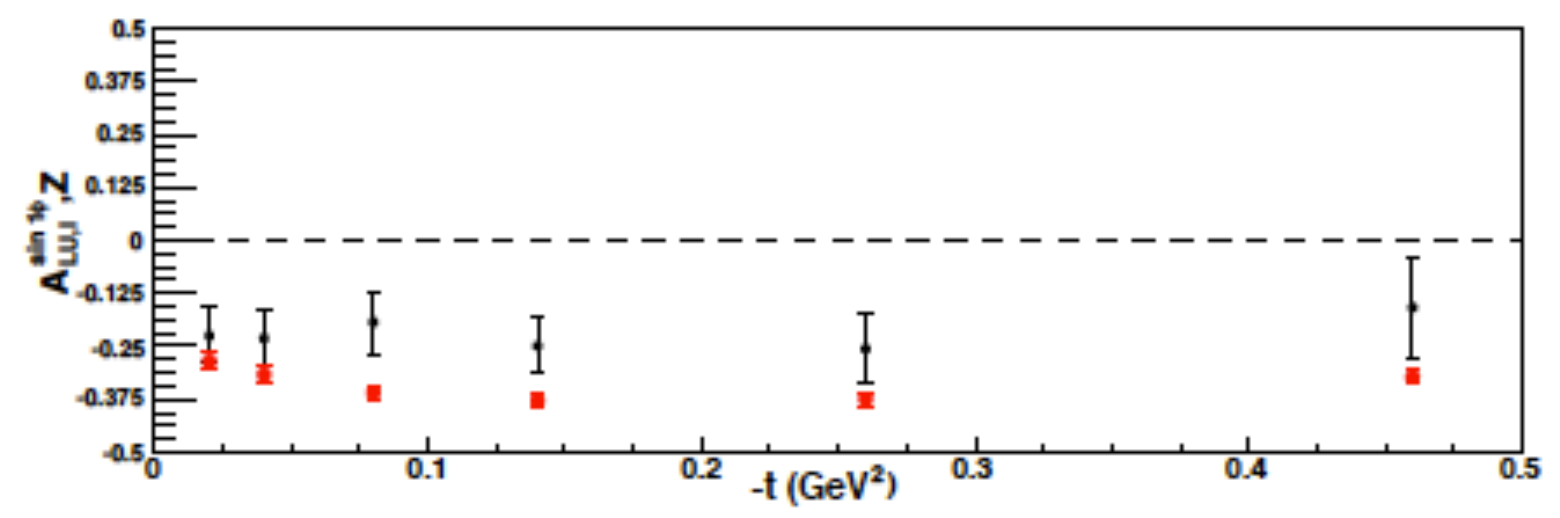}
  \end{tabular}
  \caption{Left: BCA $\cos \phi$ harmonics; Right: BSA $\sin \phi$ harmonics.}
  \label{Fig:APhi}
\end{figure}

The HERMES Collaboration released a great wealth of observables in recent years \cite{murray:2012dis}. Figure~\ref{Fig:APhi} shows the $\sin \phi$ harmonics of the Beam Spin Asymmetry (BSA), mostly sensitive to the imaginary part of $\mathcal{H}$ and the $\cos \phi$ harmonics of the Beam Charge Asymmetry (BCA), mostly sensitive to the real part of $\mathcal{H}$. The GK model is in a reasonable agreement with the data.

\subsection{JLab observables}

\begin{figure}[htb]
  \centering
%  {\rotatebox{-90}{\includegraphics[height=0.7\textwidth]{moutarde_herve_fig4}}}
  \includegraphics[width=0.7\textwidth]{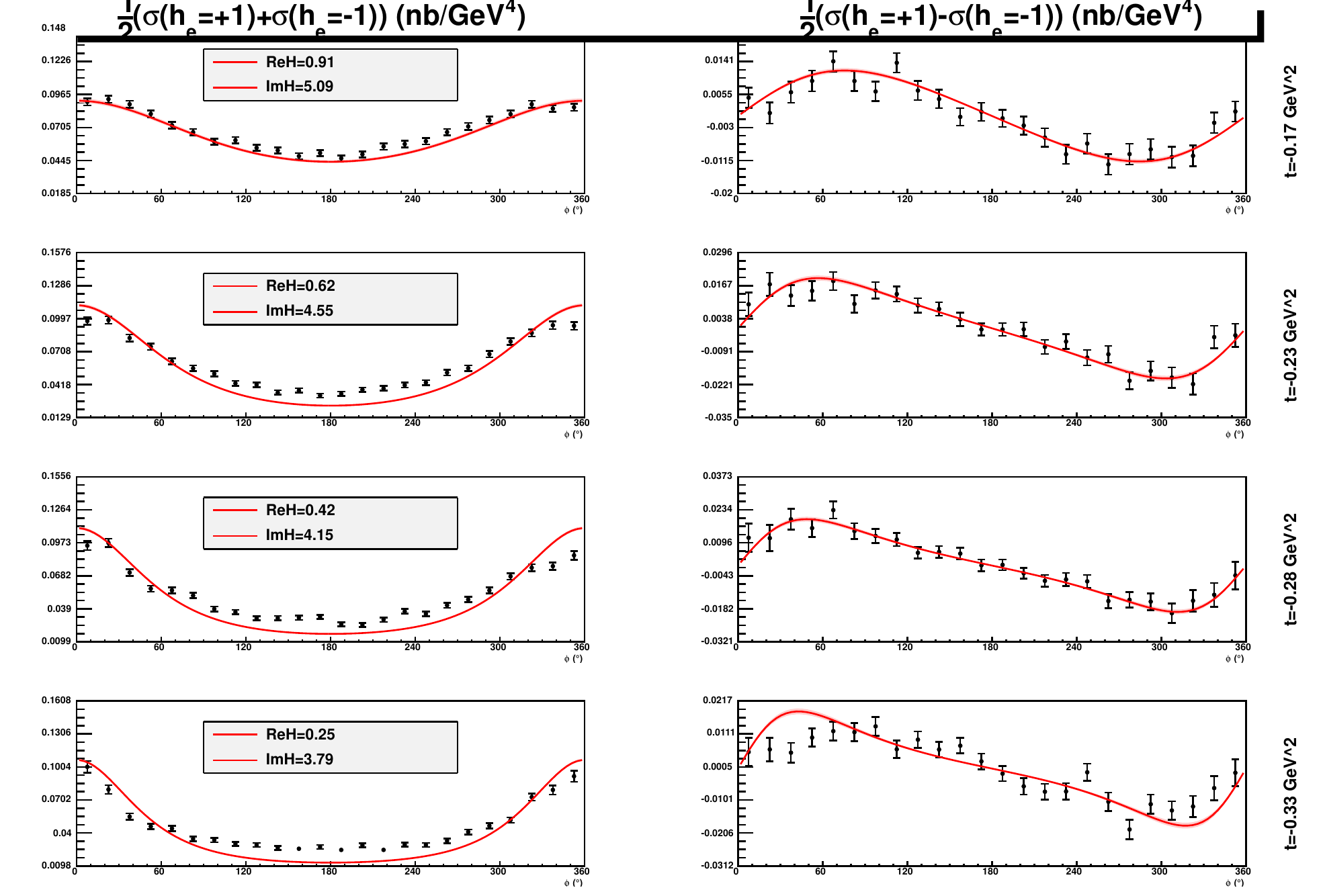}
  \caption{JLab Hall A helicity-dependent and independent cross sections.}
  \label{Fig:Sum}
\end{figure}

JLab 6~GeV DVCS observables on unpolarized targets \cite{MunozCamacho:2006hx, Girod:2007jq} cover a wide kinematic range or are highly precise. Figure~\ref{Fig:Sum} shows that the GK model tends to underestimate helicity-independent cross sections near $\phi = 180^\circ$. It also overestimates the helicity-dependent cross-sections and BSAs near $\phi = 90^\circ$, see \cite{Kroll:2012dvcs} for details.

%-------------------------------%
%--   4th part : Technical remarks   --%
%-------------------------------%

\section{Technical remarks}

\subsection{Phenomenology toolkit}

Systematic comparisons of GPD models and data require databases of experimental results and theoretical predictions, a fitting engine, tools to propagate statistic and systematic uncertainties and a flexible visualizing software. Ideally the same building blocks should be used for fits to data and designs of new experiments. Part of these building blocks are used here.

\subsection{Constraints}

After JLab's 12~GeV upgrade, phenomenologists will deal with observables with an advertised statistical accuracy of $\simeq 1~\%$. It induces some constraints on the aforementioned phenomenology toolkit. For example the evaluation of CFFs should have an accuracy better than 0.1~\% on this kinematic region, which precludes naive integration routines.

%-----------------%
%--   Conclusions   --%
%-----------------%

\section{Conclusions}

Some software components for global GPD phenomenology have been developed and extensively tested. The treatment of NLO contributions shows a surprisingly large gluon contribution in the HERMES and COMPASS kinematics, and raises the question of resummation. This study also shows how the expected accuracy of forthcoming data influences the design of software components devoted to GPD phenomenology.

%------------------------%
%--   Acknowledgements   --%
%------------------------%

\section*{Acknowledgements}

The authors would like to thank P.~Kroll, B.~Pire, L.~Szymanowski and J.~Wagner who contributed to different parts of the work presented here. The authors also thank the organizers of the 20th International Workshop on Deep-Inelastic Scattering and Related Subjects held in Bonn (March 26 - 30, 2012). This work was supported in part by the Commissariat \`a l'Energie Atomique and the GDR n$^\circ$~3034 "Chromodynamique Quantique et Physique des Hadrons".

%------------------%
%--   Bibliography   --%
%------------------%

\end{document}